\documentclass[prb,aps,twocolumn,showpacs]{revtex4-2} 
\usepackage{amsmath,amssymb,amsthm}
\usepackage{graphicx}
\usepackage{subfigure}
\usepackage{amsmath}
\usepackage{amsfonts,amssymb}
\usepackage{bm}
\usepackage{float}
\usepackage{color}
\usepackage{sidecap}
\allowdisplaybreaks
\usepackage{soul}
\setstcolor{red}
\usepackage[normalem]{ulem}
\usepackage{hyperref}
\hypersetup{colorlinks=true,breaklinks,urlcolor=blue,linkcolor=blue,citecolor=blue}

\begin{document}

\title{Evolution of spectral topology in one-dimensional long-range nonreciprocal lattices}
\author{Qi-Bo Zeng}
\email{zengqibo@cnu.edu.cn}
\affiliation{Department of Physics, Capital Normal University, Beijing 100048, China}

\author{Rong L\"u}
\affiliation{State Key Laboratory of Low-Dimensional Quantum Physics, Department of Physics, Tsinghua University, Beijing 100084, China}
\affiliation{Frontier Science Center for Quantum Information, Beijing 100084, China}

\begin{abstract}
We investigate the spectral topology of one-dimensional lattices where the nonreciprocal hoppings within the nearest $r_d$ neighboring sites are the same. For the purely off-diagonal model without onsite potentials, the energy spectrum of the lattice under periodic boundary conditions (PBCs) forms an inseparable loop that intertwines with itself in the complex energy plane and is characterized by winding numbers ranging from 1 up to $r_d$. The corresponding spectrum under open boundary conditions (OBCs), which is real in the nearest neighboring model, will ramify and take the shape of an $(r_d+1)$-pointed star with all the branches connected at zero energy. If we further introduce periodic onsite modulations, the spectrum will gradually divide into multiple separable bands as we vary the parameters. Most importantly, we find that a different kind of band gap called loop gap can exist in the PBC spectrum, separating an inner loop from an outer one with each composed by part of the spectrum. In addition, loop structures also exist in the OBC spectra of systems with onsite potentials. We further study the lattices with power-law decaying long-range nonreciprocal hopping and found that the intertwined loops in the PBC spectrum will be untangled. Finally, we propose an experimental scheme to realize the long-range nonreciprocal models by exploiting electrical circuits. Our work unveils the exotic spectral topology in the long-range nonreciprocal lattices.
\end{abstract}
\maketitle
\date{today}

\section{Introduction}\label{sect1}
The Hamiltonians of quantum systems are Hermitian in conventional quantum mechanics, yet non-Hermitian Hamiltonians have also been applied in various branches of physics. In recent years, there has been a growing interest in non-Hermitian systems for their fundamental importance as well as for their potential applications~\cite{Cao2015RMP,Konotop2016RMP,Ganainy2018NatPhy,Ashida2020AiP,Bergholtz2021RMP}. The non-Hermitian terms in the Hamiltonians may arise from the finite lifetime of quasiparticles~\cite{Fu2017arxiv,Shen2018PRL1,Yoshida2018PRB,Tao2021arxiv}, the interaction with the environment in open systems~\cite{Rotter1991RPP,Rotter2009JPA}, the complex refractive index~\cite{Musslimani2008PRL,Moiseyev2008PRL,Feng2017NatPho}, and the engineered Laplacian in electrical circuits~\cite{Schindler2011PRA,Luo2018arxiv,Lee2018ComPhy,Helbig2020NatPhys,Hofmann2020PRR,Zeng2020PRB}. Since the Hamiltonians are non-Hermitian, the eigenenergy spectra of these systems are normally complex. However, under appropriate conditions, the spectra can still be purely real, such as the $\mathcal{PT}$-symmetric~\cite{Bender1998PRL,Bender2002PRL,Bender2007RPP} and pseudo-Hermitian Hamiltonians~\cite{Mostafazadeh2002JMP,Mostafazadeh2010IJMMP,Moiseyev2011Book,Zeng2020PRB1,Kawabata2020PRR,Zeng2021arxiv}. In addition, the complex energy spectra exhibit exotic features such as exceptional point~\cite{Heiss2012JPAMT}, Weyl exceptional ring~\cite{Xu2017PRL}, point gap~\cite{Gong2018PRX}, and link or knot structures~\cite{Carlstrom2018PRA,Carlstrom2019PRB,Yang2019PRB,Yang2020PRL1,Hu2021PRL,Yang2021CPL}, which cannot be observed in Hermitian systems.  

Recently, the non-Hermitian systems with asymmetric or nonreciprocal hoppings between the lattice sites have attracted much attention~\cite{Lee2016PRL,Lieu2018PRB,Yin2018PRA}. With the presence of nonreciprocity, the system's energy spectra will be very sensitive to the changing of boundary conditions~\cite{Xiong2018JPC}. Moreover, for systems under open boundary conditions (OBCs), the bulk eigenstates are found to localize at the systems' boundaries. Such a phenomenon is called the non-Hermitian skin effect (NHSE) and has a significant effect on the systems' properties~\cite{Shen2018PRL,Yao2018PRL1,Yao2018PRL2,Alvarez2018PRB,Alvarez2018EPJ,Lee2019PRB,Zhou2019PRB,Kawabata2019PRX,Okuma2020PRB,Xiao2020NatPhys,Yoshida2020PRR,Longhi2019PRR,Yi2020PRL}. For instance, the conventional principle of bulk-boundary correspondence (BBC) of topological phases in a Hermitian system breaks down in nonreciprocal systems due to the NHSE, which has motivated several new methods to recover the BBC in non-Hermitian topological systems~\cite{Yao2018PRL1,Yao2018PRL2,Kunst2018PRL,Jin2019PRB,Yokomizo2019PRL,Herviou2019PRA,Yang2020PRL2,Zirnstein2021PRL,Zhang2022arxiv}. The nonreciprocity can also induce delocalization effects in the Anderson localization phase transition~\cite{Hatano1996PRL,Shnerb1998PRL,Gong2018PRX,Jiang2019PRB,Zeng2020PRR,Liu2021PRB1,Liu2021PRB2}. Furthermore, non-Hermiticity also introduces different types of band gap in the non-Hermitian system, e.g., the point gap and line gap in the complex energy plane~\cite{Gong2018PRX,Kawabata2019PRX}. The existence of point gap in systems under periodic boundary conditions (PBCs) is shown to be the topological origin of the skin effect in nonreciprocal systems under open boundary conditions~\cite{Borgnia2020PRL,Okuma2020PRL,Zhang2020PRL}. 

So far, most studies on non-Hermitian systems are mainly concerned with the nonreciprocal hopping between the nearest-neighboring sites. It will be interesting to ask what will happen to the non-Hermitian systems if the nonreciprocal hopping becomes long-range. More specifically, will the long-range nonreciprocity modify the spectral topology and even bring new band structures other than those observed in systems with only nearest-neighboring hoppings?   

In this paper, we answer these questions by investigating the energy spectra of one-dimensional (1D) lattices with long-range nonreciprocal hopping. The nonreciprocity is homogeneous such that the hopping amplitudes within the nearest $r_d$ neighboring sites are the same. For lattices under PBCs and OBCs, the energy spectra will exhibit entirely different structures. In the lattices without onsite modulations, the eigenenergies under PBC form an inseparable loop in the complex energy plane that intertwines with itself by crossing the zero energy $(r_d-1)$ times and another real eigenenergy $r_d$ times. The loop structure is characterized by winding numbers ranging from $1$ to $r_d$ depending on the location of base energy. However, the spectrum under OBCs, which is real in the lattices with nearest-neighboring nonreciprocal hopping, will ramify into $(r_d+1)$ branches. The OBC spectrum takes the shape of $(r_d+1)$-pointed stars with all the branches connected at zero energy and enclosed by the PBC spectra. If we further introduce periodic onsite modulations into the lattice, the PBC spectrum will divide into several separable bands gradually as the modulation gets stronger. These bands form loops and intertwine with each other, forming knot or link structures as we vary the onsite potentials. Moreover, a different type of band gap called a loop gap is found in the PBC spectra, which separates an inner loop from an outer one formed with each composed by part of the spectrum, extending the concepts of point gap and line gap in non-Hermitian systems. We also show that loop structures can exist in the OBC spectra of the long-range nonreciprocal lattices with onsite potentials. We further studied the lattices with power-law decaying long-range nonreciprocal hopping and found that the intertwined loops in the PBC spectrum will be untangled. Finally, we propose an experimental scheme to realize the long-range nonreciprocal models by using electrical circuits. Our findings reveal the exotic role of long-range nonreciprocal hopping and its subtle interplay with onsite modulations in altering the spectral topology of non-Hermitian systems.

The rest of the paper is organized as follows. In Sec.~\ref{sect2} we introduce the model Hamiltonian of the long-range nonreciprocal lattices. Then we explore the energy spectra for the systems with constant long-range nonreciprocity under both PBC and OBC in Sec.~\ref{sect3}. The influences of onsite modulations on the structures of spectra are checked in Sec.~\ref{sect4}. We further study the model with power-law decaying long-range nonreciprocal hopping in Sec.~\ref{sect5}. An experimental scheme for simulating the long-range nonreciprocal lattices is introduced in Sec.~\ref{sect6}. The last section (Sec.~\ref{sect7}) is dedicated to a summary.

\section{Model Hamiltonian}\label{sect2}
\begin{figure}[t]
  \includegraphics[width=3.3in]{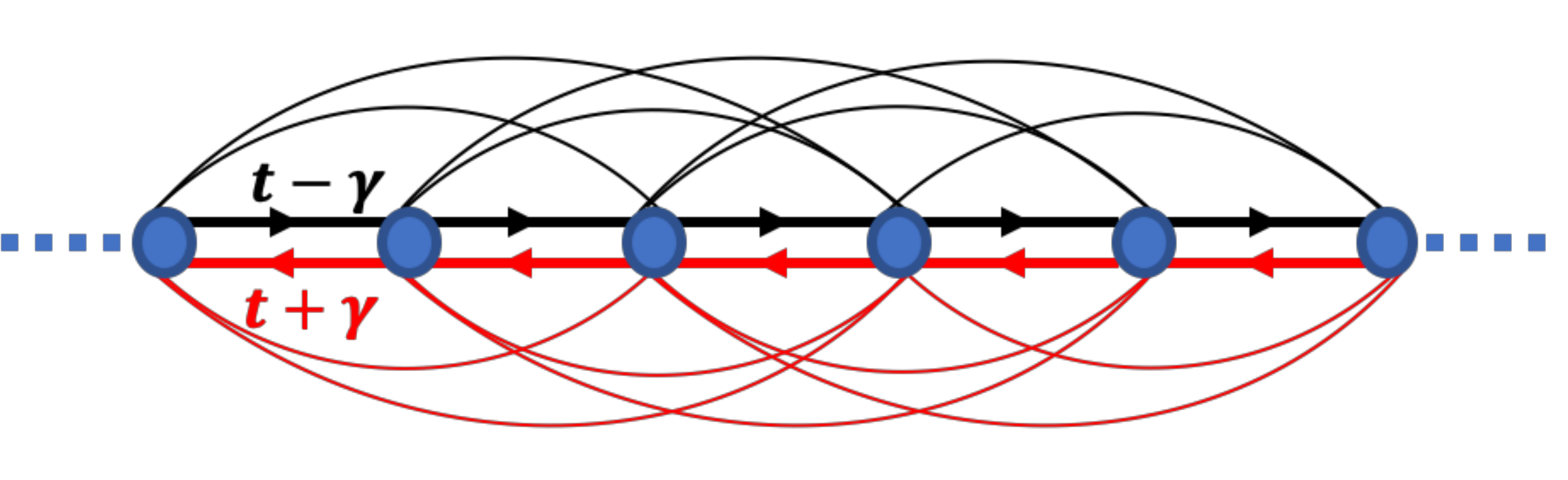}
  \caption{(Color online) Schematic illustration of the 1D lattice with long-range nonreciprocal hopping. The black and red lines represent the forward and backward hoppings within the nearest $r_d$ neighboring sites which are nonreciprocal with amplitude $(t-\gamma)$ and $(t+\gamma)$ respectively.}
\label{fig1}
\end{figure}

Figure~\ref{fig1} shows the 1D lattice with long-range nonreciprocal hopping. The black and red lines indicate the forward and backward hopping between the lattice sites, where the hopping amplitudes are $(t-\gamma)$ and $(t+\gamma)$, respectively. Such 1D long-range nonreciprocal lattices are described by the following model Hamiltonian:
\begin{equation}\label{H}
H = \sum_i V \cos (2\pi \alpha i) c_i^\dagger c_i + \sum_{1 \leq (j-i) \leq r_d} [ (t+\gamma) c_i^\dagger c_j + (t-\gamma) c_j^\dagger c_i ],
\end{equation}
where $c_i^\dagger$ ($c_i$) is the creation (annihilation) operator of a spinless fermion at the $i$th site in the lattice. $V$ is the strength of the onsite potential whose period is determined by $\alpha$. $t$ is the constant hopping amplitude and we will take $t=1$ as the energy unit throughout this paper. $\gamma$ is a real number and represents the nonreciprocal hopping. Here the particles can hop between two sites that are not farther than $r_d$, which is a positive integer and is called the cutting range. For a system with length $L$, the lattice sites are indexed as $i=1,2,\cdots,L$ under OBCs. For systems under PBCs, we set the $(L+n)$th site as $i=(L+n)mod(L)=n$. Note that we have set the lattice constant as $a=1$. 

It is known that, due to the presence of nonreciprocal hopping, the energy spectra of systems under different boundary conditions behave quite differently. By using the exact diagonalization method, we will explore the properties of energy spectra for the long-range nonreciprocal lattices with and without the presence of onsite potentials.

\section{Off-diagonal long-range nonreciprocal lattices}\label{sect3}
We first study the off-diagonal model without onsite modulations, i.e., $V=0$. The model Hamiltonian then simplifies to
\begin{equation}
H_\text{off} = \sum_{1 \leq (j-i) \leq r_d} [ (t+\gamma) c_i^\dagger c_j + (t-\gamma) c_j^\dagger c_i ].
\end{equation}
By directly diagonalizing this Hamiltonian in real space, we can obtain the energy spectrum of the system under both PBCs and OBCs. We can also transform the Hamiltonian into the momentum space and get 
\begin{equation}
\begin{split}
H_\text{off} (k) & = (t+\gamma) (e^{ik} + e^{i2k} + \cdots + e^{i r_d k}) \\
& + (t-\gamma) (e^{-ik} + e^{-i2k} + \cdots + e^{-i r_d k}).
\end{split}
\end{equation} 
Thus the energy spectrum under PBCs can also be easily obtained by calculating $H_\text{off}(k)$ with $k\in [0, 2\pi)$. When $r_d=1$, the above model reduces to the Hatano-Nelson model~\cite{Hatano1996PRL}, whose eigenenergy in momentum space is $E(k)=(t+\gamma) e^{ik} + (t-\gamma) e^{-ik}$ and forms a loop in the complex energy plane. The model we study here is a long-range generalization of the original Hatano-Nelson model. The spectrum of $H_\text{off} (k)$ also host one single band and we may expect it to form a loop. Before diving into the details of the band structure of the model, let us first check the spectrum analytically.

When $k=0$, the eigenenergy is $E_\text{off} (k=0)=2r_d t$, which is real and is represented as $E_1 = (2r_d t,0)$ in the complex energy plane. If $k \neq 0$, then the summation in $H_\text{off}(k)$ can be done directly and the eigenenergy is expressed as
\begin{widetext}
\begin{align}
E_\text{off}(k) &= (t+\gamma)e^{ik}\frac{1-e^{ir_d k}}{1-e^{ik}} + (t-\gamma)e^{-ik} \frac{1-e^{-ir_d k}}{1-e^{-ik}} \notag \\
&= t \left[ \frac{e^{ik}-e^{i(r_d+1) k}}{1-e^{ik}} + \frac{e^{-ik}-e^{-i(r_d+1) k}}{1-e^{-ik}} \right] + \gamma \left[ \frac{e^{ik}-e^{i(r_d+1) k}}{1-e^{ik}} - \frac{e^{-ik}-e^{-i(r_d+1) k}}{1-e^{-ik}} \right] \notag \\
&= t \left\lbrace -1 + \frac{\sin \left[ (r_d + \frac{1}{2}) k \right]}{\sin \left( \frac{k}{2} \right) } \right\rbrace
+ i \gamma \left\lbrace \frac{2 \sin \left[ \left( \frac{r_d+1}{2} \right) k \right] \sin \left[ \left( \frac{r_d}{2} \right) k \right] } {\sin \left( \frac{k}{2} \right) } \right\rbrace. \label{Ek} 
\end{align}
\end{widetext}
To get the real eigenenergies, we need to set the imaginary part to be zero, i.e.,
\begin{equation}
\frac{2 \sin \left[ \left( \frac{r_d+1}{2} \right) k \right] \sin \left[ \left( \frac{r_d}{2} \right) k \right] } {\sin \left( \frac{k}{2} \right)} = 0.
\end{equation}
Then we have 
\begin{align}
\sin \left[ \left( \frac{r_d}{2} \right) k \right] = 0 & \rightarrow k = \frac{2n\pi}{r_d} \quad (n=1,2,\cdots,r_d-1); \\
\sin \left[ \left( \frac{r_d+1}{2} \right) k \right] = 0 & \rightarrow k = \frac{2n\pi}{r_d+1} \quad (n=1,2,\cdots,r_d).
\end{align}
Substituting $k = \frac{2n\pi}{r_d+1}$ and $\frac{2n\pi}{r_d}$ into $E_\text{off}(k)$, we obtain two real eigenenergies whose coordinates in the complex energy plane are $E_2=(-2t,0)$ and $E_3=(0,0)$, respectively. So there are always three and only three real eigenenergies in the PBC spectrum of the off-diagonal long-range nonreciprocal lattices, and all the other eigenenergies are complex. Moreover, from Eq.~(\ref{Ek}), we have $E_\text{off}(2\pi-k)=E_\text{off}^*(k)$, thus the eigenenergies distribute symmetrically about the real axis. As $k$ sweeps from $0$ to $2\pi$, we will first encounter the first real energy $E_1 = (2r_d t,0)$ at $k=0$; then the spectrum is complex and will cross the second real energy $E_2$ at $k=\frac{2\pi}{r_d+1}$ and become complex again; when $k$ moves to $\frac{2\pi}{r_d}$, the spectrum will meet the third real energy $E_3$, i.e., the zero energy. After that, the spectrum will be complex and cross $E_2$ and $E_3$ a few more times depending on the value of $r_d$. In short, the zero energy will be crossed $(r_d-1)$ times, while the point $(-2t,0)$ will be crossed $r_d$ times. So the loop will intertwine with itself and form loops that encircle certain regions at most $r_d$ times. Next we will check these results numerically.

\begin{figure}[t]
	\includegraphics[width=3.3in]{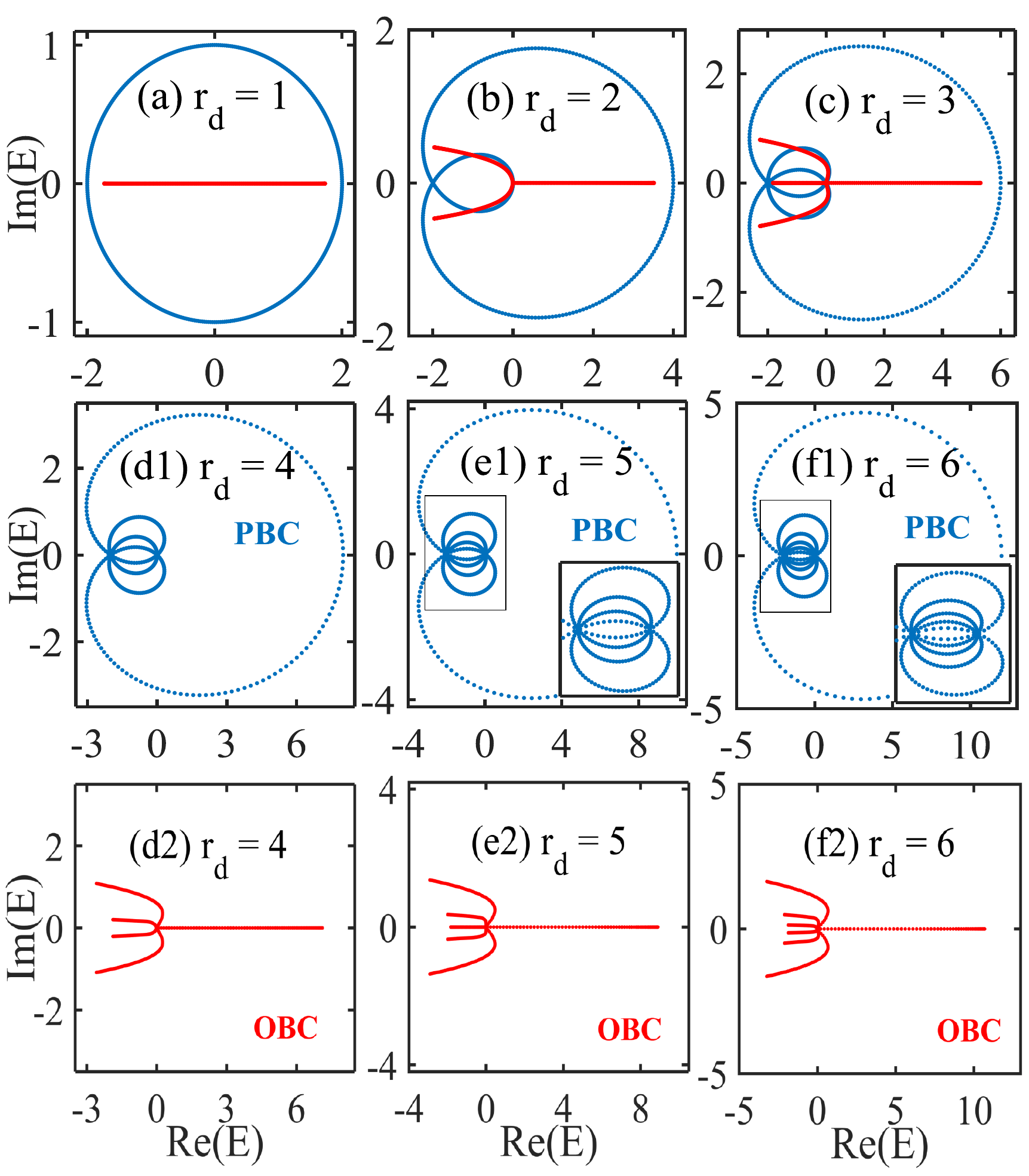}
	\caption{(Color online) Energy spectra of the 1D off-diagonal lattices with constant long-range nonreciprocal hopping. The spectrum under PBC forms loops in the complex energy plane (blue dots), while the spectrum under OBC forms a shape of an $(r_d+1)$-pointed star (red dots), with $r_d$ indicating the cutting range of the hopping amplitude. The insets in (e1) and (f1) show the enlargement of the PBC spectra enclosed by the black rectangles. The asymmetric hopping is set to be $\gamma=0.5$ and the lattice size is $L=400$.}
	\label{fig2}
\end{figure} 

\begin{figure}[t]
  \includegraphics[width=3.3in]{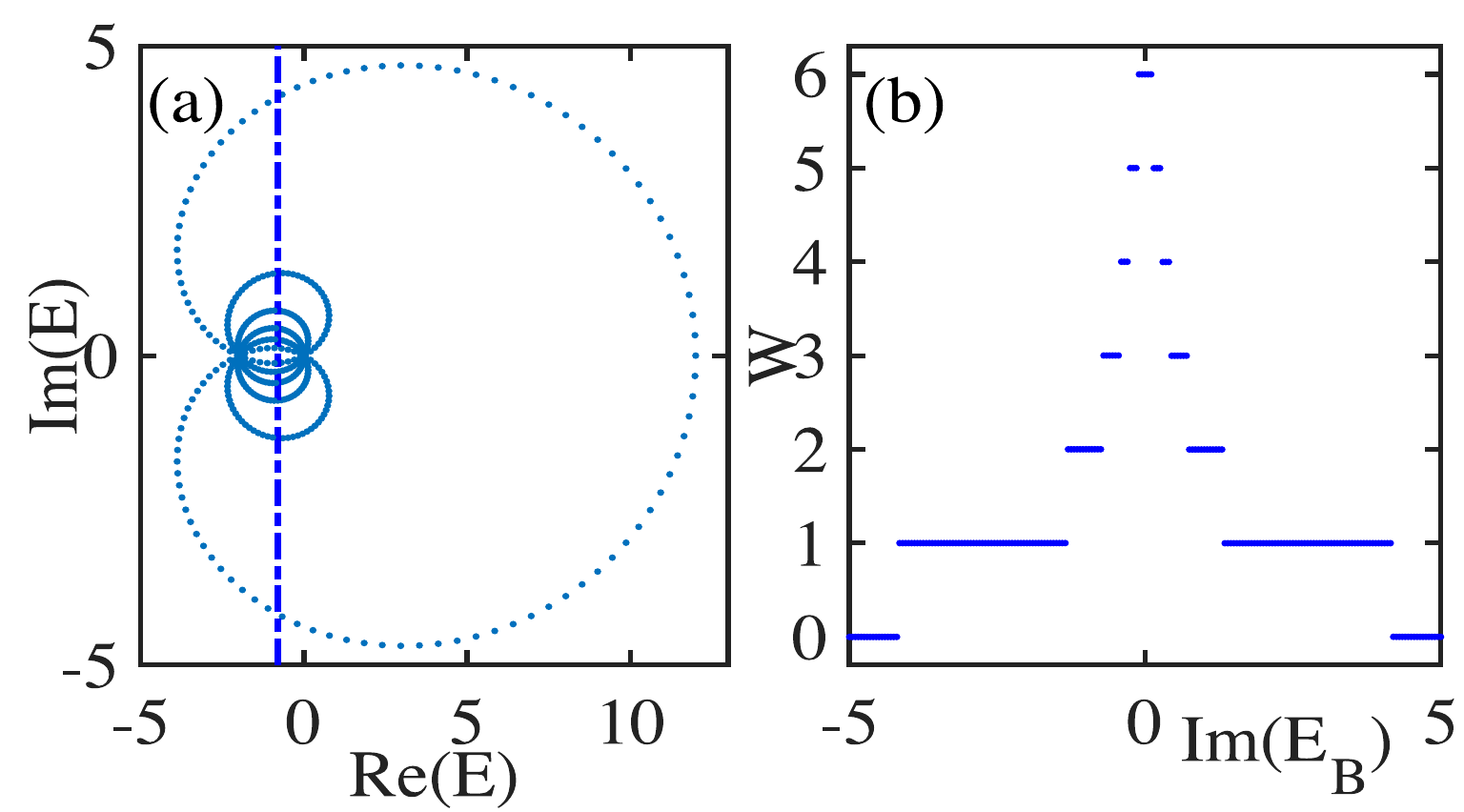}
  \caption{(Color online) (a) The PBC spectra of the off-diagonal long-range nonreciprocal lattices with $r_d=6$. The blue dashed line indicates the points with the real part being $-0.8$ and imaginary parts between $[-5,5]$. (b) The winding number calculated for different base energies $E_B$ along the blue dashed line in (a). Other parameters are the same as in Fig.~\ref{fig2}.}
\label{fig3}
\end{figure}

In Fig.~\ref{fig2}, we plot the energy spectra for the off-diagonal nonreciprocal lattices with different cutting ranges $r_d$. The blue and red dots represent the spectra under PBCs and OBCs, respectively. For the lattices with asymmetric hopping present only between the nearest-neighboring sites, i.e., $r_d=1$, the spectrum of the system is complex and forms a closed loop in the complex energy plane, where the real parts of the spectrum are confined in the regime $[-2t,2t]$; see Fig.~\ref{fig2}(a). The corresponding OBC spectrum is purely real as indicated by the red dots, which are fully enclosed by the loop formed by the PBC spectrum. This phenomenon has been well studied and the changing of the spectral topology is connected to the non-Hermitian skin effect~\cite{Borgnia2020PRL,Okuma2020PRL,Zhang2020PRL}. If the nonreciprocal hopping becomes long range, we find that the PBC spectra of the systems will intertwine with themselves and form smaller loops that are connected to the outside larger loops at the point $(-2t,0)$ and the zero energy, as shown in Fig.~\ref{fig2}(b)-(f). The largest eigenenergy that is real is $2r_d t$, which corresponds to the energy at $k=0$. In addition, we can see that the loop cross the zero energy and the point $(-2t,0)$ by $(r_d-1)$ and $r_d$ times, consistent with our analytic results obtained above. When the cutting range changes from $r_d$ to $(r_d+1)$, the loop will intertwine one more time with itself by crossing these two points, which forms a new small loop and changes the spectral topology of the spectrum. These loop structures in the complex energy plane can be characterized by winding numbers defined as follows
\begin{equation}\label{W}
W = \frac{1}{2\pi i} \int_0^{2\pi} dk \partial_k \text{arg} [ E(k) - E_B],
\end{equation}
where $E_B$ is the base energy. Depending on the location of base energy $E_B$, the winding number will change from $0$ to $r_d$. For example, in Fig.~\ref{fig3}, we show the winding numbers for the case with $r_d=6$. Choosing the base energies along the blue dashed line in Fig.~\ref{fig3}(a) with the real part being $-0.8$ and imaginary parts within $[-5,5]$, the winding number changes from $0$ to $6$ stepwise, as shown in Fig.~\ref{fig3}(b). When the base energy is located inside the innermost loop of the spectra, $W$ is the largest. If $E_B$ is located outside the PBC spectrum, we have $W=0$. Similar phenomena can also be observed in lattices with other values of $r_d$. So the introduction of long-range nonreciprocal hopping results in much richer topological structures in the PBC spectra of non-Hermitian systems.

The loop structures in the PBC spectra of the long-range nonreciprocal lattices are quite different from the loop structures in the systems with only nearest-neighboring asymmetric hopping. The introduction of long-range hopping can result in richer phenomena in the spectrum, in both the Hermitian and non-Hermitian systems. For instance, the topological edge modes are investigated in the Hermitian Su-Schrieffer-Heeger (SSH) model with long-range hopping, where winding numbers can be defined to characterize the topologically nontrivial phase~\cite{Dias2022PRB}. Notice that the winding number defined in this work is different from the one defined in Ref.~\cite{Dias2022PRB}. The winding number defined in Eq.~\ref{W} characterizes the loop structures in the PBC spectra.

\begin{figure}[t]
  \includegraphics[width=3.3in]{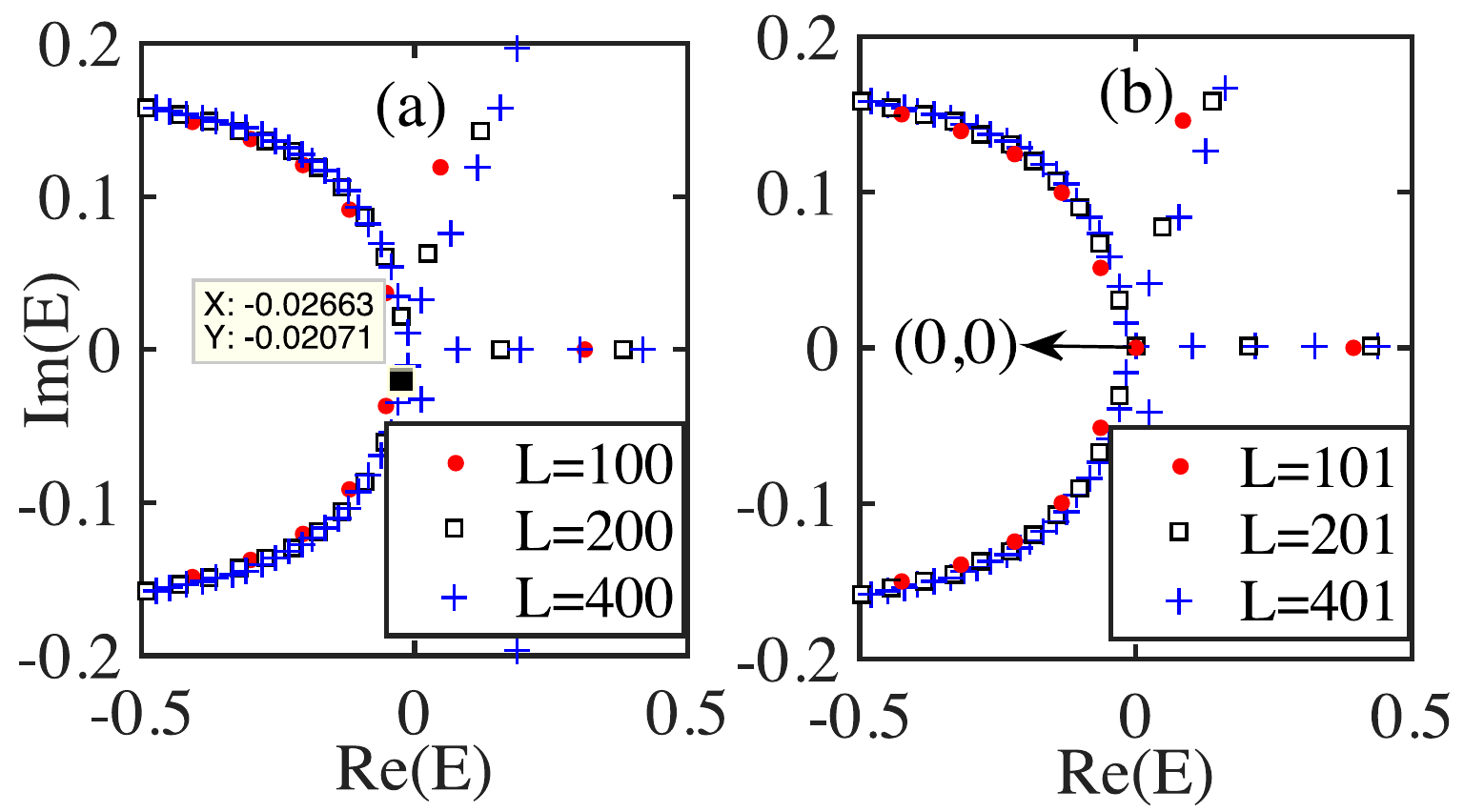}
  \caption{(Color online) The OBC spectra of the off-diagonal long-range nonreciprocal lattices with $r_d=4$. (a) and (b) show the enlargements of the spectra with different lattice sizes. The $X$ and $Y$ numbers in (a) indicate the coordinate of the black solid square, which is the energy closest to zero for the lattice with $L=200$.}
\label{fig4}
\end{figure}

Now we turn to the OBC spectra of the long-range systems. As indicated by the red dots in Fig. \ref{fig2}, the spectra of the systems under OBCs are always enclosed by the PBC spectra. If we only consider the nearest neighboring hopping, then the OBC spectrum is purely real, as shown in Fig. \ref{fig2}(a) for the case with $r_d=1$. When $r_d$ becomes larger, the real OBC spectrum will ramify into several branches at one point, where the branches are either reside on the real axis or are distributed symmetrically about the real energy axis. Thus parts of the OBC spectra are real while the others are complex conjugate pairs; see Fig. \ref{fig2}(b)-(f). From these numerical results, we can find that the spectra form a shape of $(r_d+1)$-pointed stars. Moreover, all the $(r_d+1)$ branches of the stars are connected at zero energy. To better illustrate this, we enlarge the OBC spectra for the lattices with $r_d=4$ with different lattice sizes, as shown in Fig.~\ref{fig4}. We find that when $L$ is even, the connecting point of the five branches of the spectrum will get closer to zero as $L$ becomes larger. For instance, when $L=200$, the energy closest to zero is around $(\pm0.026,\pm0.02)$, as indicated by the black solid square in Fig.~\ref{fig4}(a). When $L$ increases to $400$, the point moves further toward zero, as represented by the blue cross. However, if $L$ is odd, then the connecting point is always located at zero energy, see Fig.~\ref{fig4}(b). So, in the limit $L\rightarrow \infty$, the connecting point will be the zero energy point. The phenomenon is also observed in other cases with different cutting ranges of the long-range nonreciprocal hopping. We thus conclude that the OBC spectra for the long-range nonreciprocal lattices will form $(r_d+1)$-pointed stars in the complex energy plane with all the branches connected at zero energy. 

\section{Long-range nonreciprocal lattices with onsite modulation}\label{sect4}
The above discussions have been focused on off-diagonal models where onsite modulations are absent. In the nonreciprocal lattices with only nearest-neighboring asymmetric hopping, the interplay between the onsite modulation and nonreciprocal hopping can lead to quite interesting phenomena. For instance, in the 1D lattices with quasiperiodic or random onsite modulations, the presence of nonreciprocal hopping can induce a delocalization effect~\cite{Gong2018PRX,Jiang2019PRB,Zeng2020PRR,Liu2021PRB1,Liu2021PRB2}. Moreover, exotic loop structures consisting of both localized and extended states have been found in such lattices~\cite{Zeng2020PRR}. Regarding the long-range nonreciprocal lattices studied in this work, it will be interesting to study how the spectral topology of the system will evolve under the influence of the onsite modulations. 

\begin{figure}[t]
  \includegraphics[width=3.3in]{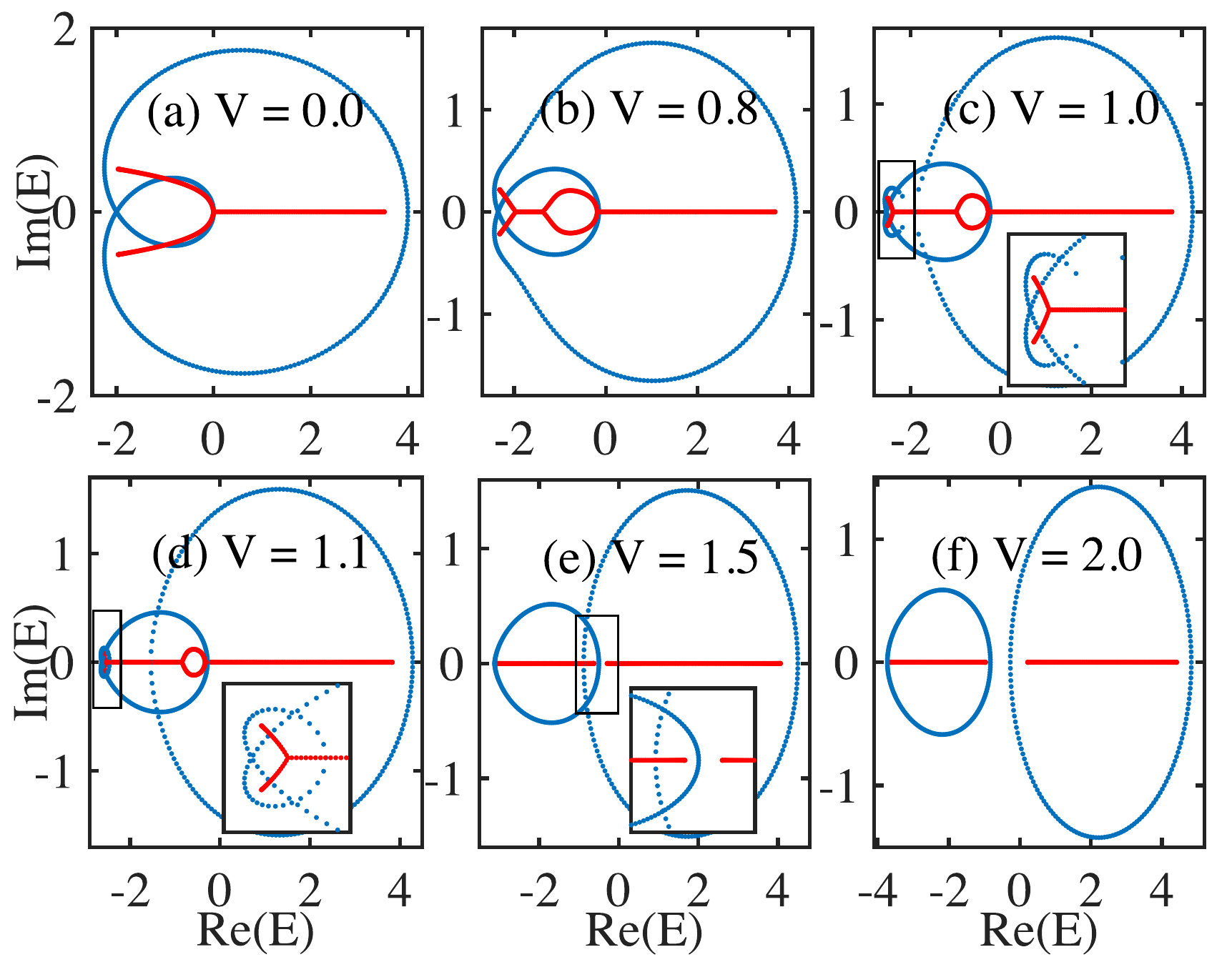}
  \caption{(Color online) Energy spectra of the 1D long-range commensurate nonreciprocal lattices with $\alpha=1/2$ and different commensurate onsite potentials $V$. The blue dots indicate the PBC spectra while the red dots indicate the OBC spectra. Here we have chosen $\gamma=0.5$, $r_d=2$, and $L=400$.}
\label{fig5}
\end{figure}

Here we consider the long-range nonreciprocal lattices with commensurate onsite potentials. In Eq.~(\ref{H}), we take $V \neq 0$ and set $\alpha = p/q$ with $p$ and $q$ being coprime integers. The period of the onsite modulation is $q$, and the energy spectrum of the system will divide into multiple bands in general. In Fig.~\ref{fig5}, we present the PBC as well as the OBC energy spectra for the long-range model with $r_d=2$ and $\alpha=1/2$. For the system under PBCs, we find that when the onsite modulation is weak, the spectrum is still an inseparable loop intertwining with itself, similar to the case without onsite potentials, as compared in Figs.~\ref{fig5}(a) and \ref{fig5}(a)(b). Now we increase the value of $V$, and the spectrum gradually transforms into two loops that are intertwined with each other and form a link structure. When the onsite potential becomes even stronger, the two loops will finally be separated and the spectrum exhibits a two-band structure. On the other hand, for the OBC spectra (red dots in Fig.~\ref{fig5}), the band structures also exhibit a series of interesting transformations as we tune the onsite potentials. In systems with small $V$, the OBC spectrum is inseparable and parts of the spectrum are purely real while others bifurcate into complex conjugate pairs. Interestingly, we find loop structures in the OBC spectrum, as shown in Figs.~\ref{fig5}(b)-~\ref{fig5}(d). It has been reported that loop structures can be observed in the OBC spectra of nonreciprocal quasiperiodic lattices in Ref.~\cite{Zeng2020PRR}. Here we show that a similar phenomenon also exists in the OBC spectra of the long-range nonreciprocal lattices with commensurate onsite potentials. As $V$ gets stronger, the size of the loop, as well as the bifurcation in the OBC spectrum, will decrease, disappearing in the end. The spectrum will also divide into two separable bands and become entirely real when the onsite potential is strong enough.   

\begin{figure}[t]
  \includegraphics[width=3.3in]{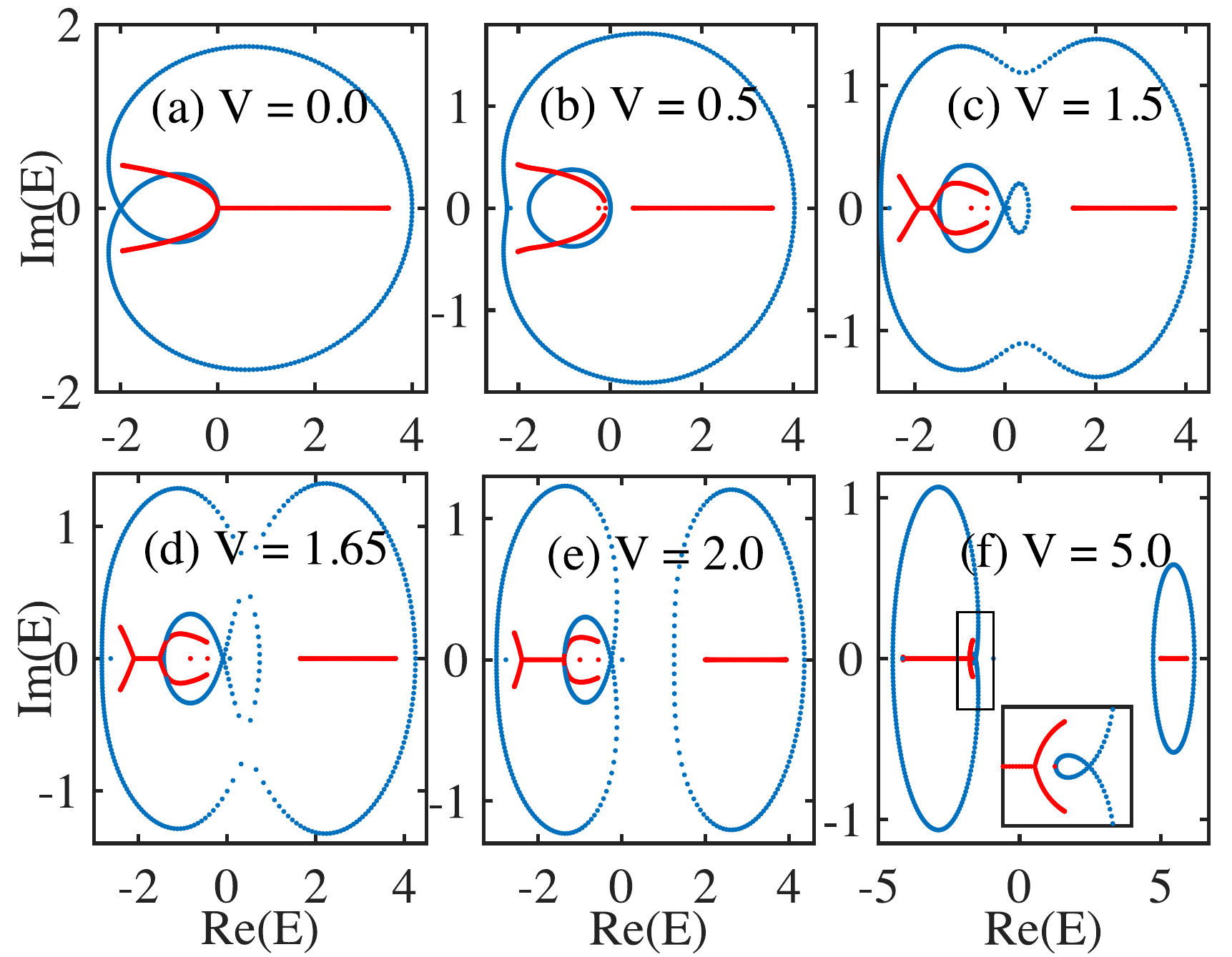}
  \caption{(Color online) The energy spectra of the 1D long-range commensurate nonreciprocal lattices with $\alpha=1/3$ and different commensurate onsite potentials $V$. The blue dots indicate the PBC spectra while the red dots indicate the OBC spectra. Other parameters are $\gamma=0.5$, $r_d=2$, and $L=400$.}
\label{fig6}
\end{figure}

If we change the period of the onsite modulation, more interesting topological structures might emerge in the complex energy spectra of the long-range nonreciprocal systems. Figure~\ref{fig6} shows the energy spectra for the long-range nonreciprocal lattices with $r_d=2$ and $\alpha=1/3$. The PBC spectra exhibit very different structures compared with the case shown in Fig.~\ref{fig5} with $\alpha=1/2$. More specifically, we observe the structures with one smaller loop residing inside another larger loop; see Fig.~\ref{fig6}(b). The PBC spectrum is divided into two separable loops and they are gapped in the complex energy plane, thus the gap can be called a loop gap. Such band structures are different from the point gap and line gap reported in previous studies on non-Hermitian systems. The spectrum can also be described by the winding number defined in Eq.(\ref{W}). If the base energy $E_B$ locates inside the smaller inner loop, the winding number is $2$. If $E_B$ moves out of the inner loop but remains inside the outer loop, the winding number will jump from $2$ to $1$. In fact, the whole region enclosed by the inner loop is characterized by $W=2$ while the gapped region between the inner and outer loop is characterized by $W=1$. The winding number for the region outside the outer loop is $0$. Differently from the inseparable loop in the model without onsite potentials in the previous section, where the loop intertwines with itself and all the loops are connected, the inner and outer loops in Fig.~\ref{fig6}(b) are separated by the loop gap. Recently, the loop gap was reported in interacting non-Hermitian systems~\cite{Shen2021arxiv}. Here we show that the loop gap structures can exist in our noninteracting long-range nonreciprocal systems. Differently from the loop gap in Ref.~\cite{Shen2021arxiv}, the inner and outer loops in our model are not concentric. Besides, we can see that the inner loop can host more complex structures than a simple loop, as shown in Fig.~\ref{fig6}(c), where two smaller loops  are connected inside the outer loop. Furthermore, as the onsite modulation becomes stronger, the outer loop will merge with the inner loop. The two loops thus become inseparable and the loop gap disappears; see Fig.~\ref{fig6}(d). If $V$ further increases, the spectrum divides into two parts that are now separated by a line gap [Figs.~\ref{fig6}(e) and \ref{fig6}(f)].

From the above discussions, we find that the interplay between the long-range nonreciprocal hopping and the onsite modulations can result in exotic spectral topology that cannot be observed in the non-Hermitian lattices with only nearest neighboring hopping. The existence of long-range hopping can significantly modify the topology of the spectrum in the complex energy plane and even lead to the emergence of a loop gap, which is a kind of band gap different from the point gap and line gap in non-Hermitian systems. We thus provide a method to study the band theory in non-Hermitian systems by tuning the cutting range of the long-range nonreciprocal hopping and the strength of onsite modulations.

\section{Long-range lattices with power-law decaying nonreciprocal hopping}\label{sect5}
\begin{figure}[t]
	\includegraphics[width=3.3in]{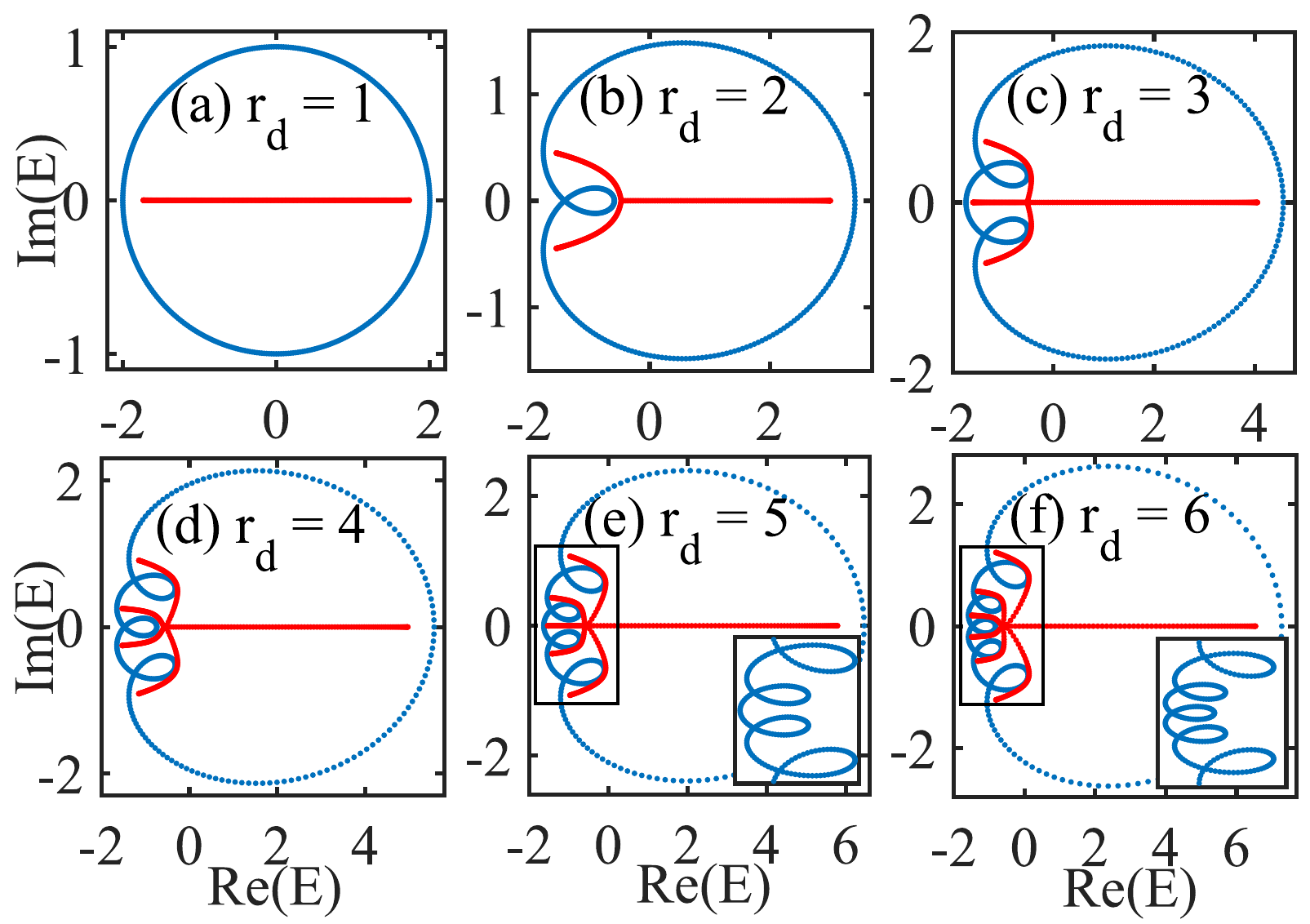}
	\caption{(Color online) Energy spectra of the 1D off-diagonal lattices with power-law decaying nonreciprocal hopping which scales as $1/|i-j|^a$. The blue dots indicate the PBC spectra while the red dots represent the OBC spectra. $r_d$ is the cutting range of the hopping amplitude. The insets in (e) and (f) show the enlargement of the PBC spectra enclosed by the black rectangles. Other parameters:$\gamma=0.5$, $a=0.5$, and $L=400$.}
	\label{fig7}
\end{figure}

If the long-range nonreciprocal hopping is not constant, but decays with the distance of the lattice sites, the energy spectrum of the system would exhibit different features. For instance, we consider the long-range nonreciprocal lattices with power-law decay in the hopping amplitude, whose model Hamiltonian is expressed as 
\begin{equation}
	H_1 = \sum_{1 \leq (j-i) \leq r_d} \left[ \frac{t+\gamma}{|i-j|^a} c_i^\dagger c_j + \frac{t-\gamma}{|i-j|^a} c_j^\dagger c_i \right].
\end{equation}
The hopping amplitudes between the nearest $r_d$ sites decay as $1/r^a$ with $r=|i-j|$ being the distance of the lattice sites. The power exponent $a$ is a positive real number. In Fig.~\ref{fig7}, we present the OBC (red dots) and PBC (blue dots) spectra of the lattice with $\gamma=0.5$ and $a=0.5$ as the cutting range $r_d$ changes. Similarly to the model with constant long-range nonreciprocal hopping, the spectra of the lattice with power-law decaying hopping under open boundary also shows an $(r_d+1)$-pointed star structure when $r_d>1$, except that the branches are no longer connected at zero energy. On the other hand, the PBC spectra still form loops in the complex energy plane, but they are not intertwined with each other. We can find  $(r_d-1)$ small loops in the PBC spectrum that are separable from each other, which is different from the spectrum of the model with constant long-range nonreciprocal hopping shown in Fig.~\ref{fig2}. 

\begin{figure}[t]
	\includegraphics[width=3.3in]{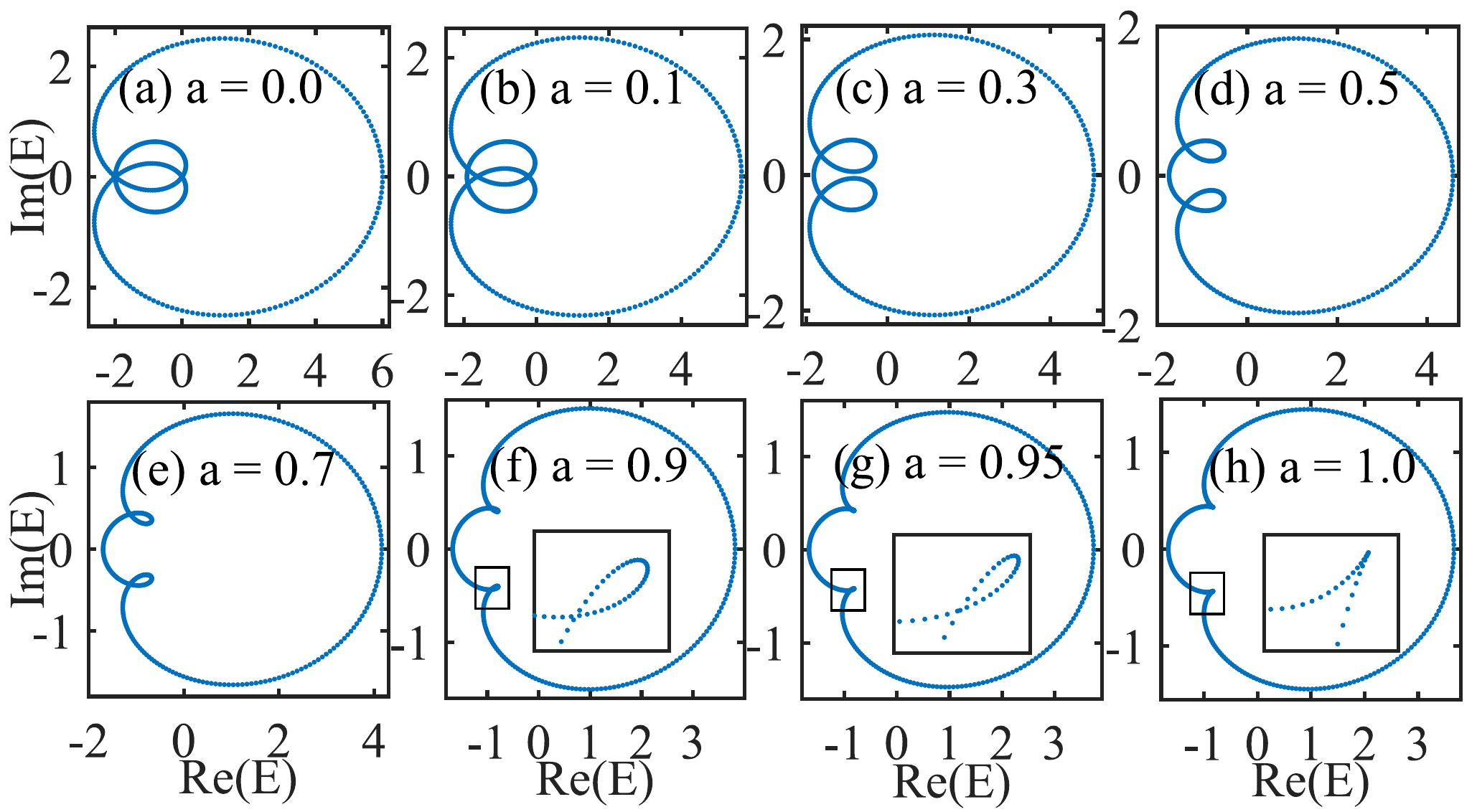}
	\caption{(Color online) The variation of the PBC spectra of the 1D lattice with power-law decaying nonreciprocal hopping as the power exponent $a$ changes. The insets in (f)-(h) show the zooming in of the region enclosed by the black rectangles. Other parameters: $\gamma=0.5$, $r_d=3$.}
	\label{fig8}
\end{figure}

To further unravel the influence of the power-law decaying on the PBC spectra of the nonreciprocal systems, we calculate the energy spectra of the system with $r_d=3$, $\gamma=0.5$, and different power exponents $a$, as shown in Fig.~\ref{fig8}. When the nonreciproctity is constant, i.e., $a=0$, the PBC spectra will form loops that intertwine with each other; see Fig.~\ref{fig8}(a). When the hopping becomes power-law decayed, the two small loops will gradually separate as $a$ gets larger and become untangled eventually. Meanwhile, the two small loops also shrink and disappear when $a=1$. So when the long-range nonreciprocal hopping decays faster, which corresponds to larger $a$, the loop structure in the PBC spectrum will become untangled. The small loops will disappear in the end and the PBC spectrum becomes a single big loop in the complex energy plane. This implies that the decay in the long-range nonreciprocal hopping could change the spectral topology of non-Hermitian systems in a significant way.

\section{Experimental realization in electrical circuits}\label{sect6}
\begin{figure}[t]
	\includegraphics[width=3.3in]{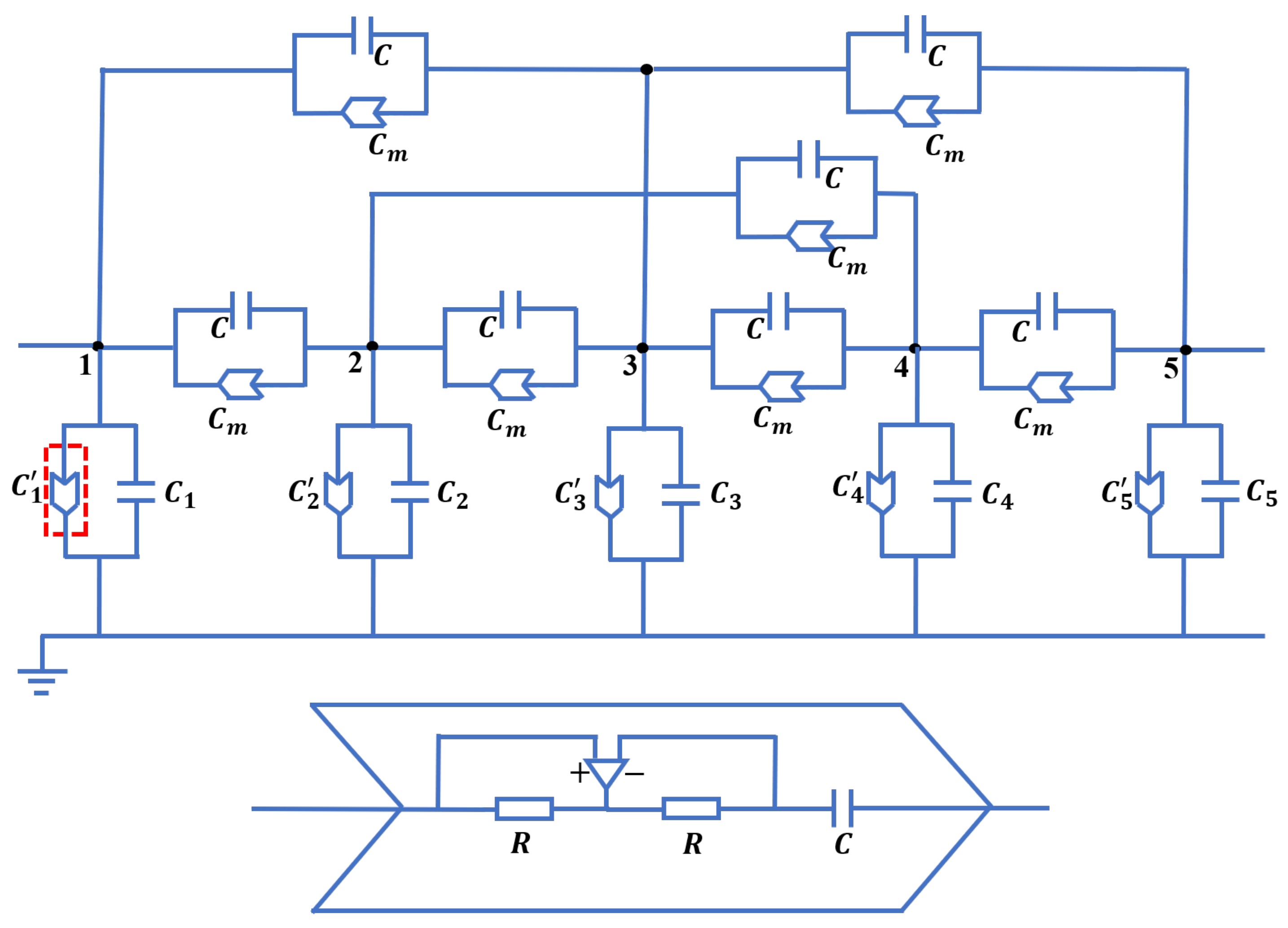}
	\caption{(Color online) The schematic of an electrical circuit for realizing the lattice model with constant nonreciprocal hopping amplitudes between the nearest and next nearest neighboring sites. The circuit is composed of capacitance and a negative impedance converter with current inversion (INIC) (the device enclosed by the red rectangle). The black dots labeled with numbers are the nodes in the circuits, which correspond to the lattices sites in the model. The lower panel shows the structure of the INIC, whose impedance depends on the direction of the current running through it.}
	\label{fig9}
\end{figure}

To experimentally verify the properties of the long-range nonreciprocal lattices, we can use electrical circuits to simulate the models. Electrical circuits have been demonstrated to be a powerful platform in simulating non-Hermitian systems~\cite{Imhof2018NatPhys,Hofmann2019PRL,Helbig2019arxiv}. The electrical circuit for the long-range nonreciprocal lattices with $r_d=2$ is shown in Fig.~\ref{fig9}. The numbers indicate the nodes in the circuit where
external sources or measuring instruments can be linked. The circuit consists of capacitors and a negative impedance converter with current inversion (INIC)~\cite{Hofmann2019PRL,Chen2009}, which is used to simulate the asymmetric hopping in the lattice. The structure of the INIC is presented in the lower panel of Fig.~\ref{fig9}, which is composed of two resistors, one capacitor, and one operational amplifier. The capacitance of the INIC depends on the orientation of the current running through the device. If the current runs from left to right, the capacitance is negative, while, if the current reverses, the capacitance becomes positive. For a node $j$ in the circuit, suppose that the input current and electrical potential at that node are $I_j$ and $V_j$, respectively. Then from Kirchhoff's law we have
\begin{equation}
	I_j = \sum_i Y_{ji} (V_j - V_i) + X_j V_j,
\end{equation}
where $j$ represents all other nodes linked to node $j$ with conductance $Y_{ji}$. $X_j$ is the conductance of node $j$. Taking all the nodes in the circuit into consideration, we can write the voltage-current relation into a compact matrix form as 
\begin{equation}
	\boldsymbol{I} = \boldsymbol{JV},
\end{equation}
where $\boldsymbol{J}$ is the Laplacian of the circuit. The circuit Laplacian can be used to model the tight-binding lattices Hamiltonian. 

\begin{widetext}
For the electrical circuit shown in Fig.~\ref{fig9}, the circuit Laplacian is
\begin{equation}
	\boldsymbol{J}=i\omega 
	\left(
	\begin{array}{ccccc}
		C_1+2C+2C_m-C_1^\prime & -(C+C_m) & -(C+C_m) & 0 & \cdots \\
		-(C-C_m) & C_2+3C+C_m-C_2^\prime & -(C+C_m) & -(C+C_m) & \cdots \\
		-(C-C_m) & -(C-C_m) & C_3+3C+C_m-C_3^\prime & -(C+C_m) & \cdots \\
		\vdots & \vdots & \vdots & \ddots & \vdots \\
		0 & 0 & 0 & \cdots & C_L+2C-2C_m-C_L^\prime
	\end{array}
	\right).
\end{equation}
This matrix corresponds to the long-range lattice with constant nonreciprocal hopping with $r_d=2$ and under open boundary conditions. If we want to simulate the model with onsite modulations, we can set 
\begin{equation}
	 C_1^\prime=2(C+C_m), \qquad C_L^\prime=2(C-C_m), \qquad C_j^\prime=3C+C_m \quad for \quad 1<j<L.
\end{equation}
Then the $C_j$ left in the diagonal terms represent the onsite modulations. If we want to study the model without onsite potential, we may further set
\begin{equation}
	C_j = 0 \quad for \quad 1\leq j \leq  L,
\end{equation}
and the circuit simulates a purely off-diagonal lattice model. The electrical circuits for the model under periodic boundary conditions and for the model with power-decaying nonreciprocal hopping can be constructed similarly. The energy spectrum is obtained from the admittance spectrum of the circuit and can be used to detect the spectral properties reported in this work.
\end{widetext}

\section{Summary}\label{sect7}
In summary, we have studied the energy spectra of one-dimensional lattices with homogeneous nonreciprocal hopping existing in the nearest $r_d$ neighboring sites. We show that, in the off-diagonal models without onsite potentials, the PBC spectra always form loops that are characterized by winding numbers up to $r_d$. However, the corresponding OBC spectra will ramify and take the shape of $(r_d+1)$-pointed stars with all the branches connected at zero energy. By further introducing onsite modulations into the long-range lattices, we find that the energy spectra are gradually divided into different bands as the onsite potentials grow stronger. Besides, loop structures are also found in the OBC spectra. We also report the existence of loop gaps in this long-range nonreciprocal lattice with the presence of onsite modulations, which separates an inner loop from an outer one in the PBC spectra and features a different kind of energy gap in non-Hermitian systems. The results illustrate that by tuning the long-range nonreciprocity and the onsite modulations, the energy spectra could evolve into various exotic band structures in the complex energy plane. We further study the spectral properties of 1D lattices with power-law decaying nonreciprocal hopping. Finally, we propose a practical scheme by using electrical circuits to simulate the long-range nonreciprocal systems. Our work unveils the effect of long-range nonreciprocal hopping on the energy spectra and provides a method to obtain diverse spectral topology in non-Hermitian systems. 

\section*{Acknowledgments}
This work is Supported by Open Research Fund Program of the State Key Laboratory of Low-Dimensional Quantum Physics. Q.-B.-Z. would like to thank Y. Xu for helpful discussions. R.-L is supported by the NSFC under Grant No. 11874234 and the National Key Research and Development Program of China (2018YFA0306504).

%


\begin{thebibliography}{}
\bibitem{Cao2015RMP}{H. Cao and J. Wiersig, Rev. Mod. Phys. \textbf{87,} 61 (2015).}

\bibitem{Konotop2016RMP}{V. V. Konotop, J. Yang, and D. A. Zezyulin, Rev. Mod. Phys. \textbf{88,} 035002 (2016).}

\bibitem{Ganainy2018NatPhy}{R. El-Ganainy, K. G. Makris, M. Khajavikhan, Z. H. Musslimani, S. Rotter, and D. N. Christodoulides, Nat. Phys. \textbf{14,} 11 (2018).}

\bibitem{Ashida2020AiP}{Y. Ashida, Z. Gong, and M. Ueda, Advances in Physics \textbf{69,} 3 (2020).}

\bibitem{Bergholtz2021RMP}{E. J. Bergholtz, J. C. Budich, and F. K. Kunst, Rev. Mod. Phys. \textbf{93,} 015005 (2021).}

\bibitem{Fu2017arxiv}{V. Kozii and L. Fu, arXiv:1708.05841 (2017).}

\bibitem{Shen2018PRL1}{H. Shen and L. Fu, Phys. Rev. Lett. \textbf{121,} 026403 (2018).}

\bibitem{Yoshida2018PRB}{T. Yoshida, R. Peters, and N. Kawakami, Phys. Rev. B \textbf{98,} 035141 (2018)}

\bibitem{Tao2021arxiv}{Y.-L. Tao, T. Qin, and Y. Xu, arXiv:2111.03348 (2021).}

\bibitem{Rotter1991RPP}{I. Rotter, Rep. Prog. Phys. \textbf{54,} 635 (1991).}

\bibitem{Rotter2009JPA}{I. Rotter, J. Phys. A \textbf{42,} 153001 (2009).}

\bibitem{Musslimani2008PRL}{Z. H. Musslimani, K. G. Makris, R. El-Ganainy, and D. N. Christodoulides, Phys. Rev. Lett. \textbf{100,} 030402 (2008).}

\bibitem{Moiseyev2008PRL}{S. Klaiman, U. G\"unther, and N. Moiseyev, Phys. Rev. Lett. \textbf{101,} 080402 (2008).}

\bibitem{Feng2017NatPho}{L. Feng, R. El-Ganainy, and L. Ge, Nat. Photonics \textbf{11,} 752 (2017).}

\bibitem{Schindler2011PRA}{J. Schindler, A. Li, M. C. Zheng, F. M. Ellis, and T. Kottos, Phys. Rev. A \textbf{84,} 040101(R) (2011).}

\bibitem{Luo2018arxiv}{K. F. Luo, J. J. Feng, Y. X. Zhao, and R. Yu, arXiv:1810.09231 (2018).}

\bibitem{Lee2018ComPhy}{C. H. Lee, S. Imhof, C. Berger, F. Bayer, J. Brehm, L. W. Molenkamp, T. Kiessling, and R. Thomale, Commun. Phys. \textbf{1,} 39 (2018).}

\bibitem{Helbig2020NatPhys}{T. Helbig, T. Hofmann, S. Imhof, M. Abdelghany, T. Kiessling, L. W. Molenkamp, C. H. Lee, A. Szameit, M. Greiter, and R. Thomale, Nat. Phys. \textbf{16,} 747 (2020).}

\bibitem{Hofmann2020PRR}{T. Hofmann, T. Helbig, F. Schindler, N. Salgo, M. Brzezińska, M. Greiter, T. Kiessling, D. Wolf, A. Vollhardt, A. Kabaši, C. H. Lee, A. Bilušić, R. Thomale, and T. Neupert, Phys. Rev. Research \textbf{2,} 023265 (2020).}

\bibitem{Zeng2020PRB}{Q.-B. Zeng, Y.-B. Yang, and Y. Xu, Phys. Rev. B \textbf{101,} 020201(R) (2020).}

\bibitem{Bender1998PRL}{C. M. Bender and S. Boettcher, Phys. Rev. Lett. \textbf{80,} 5243 (1998).}

\bibitem{Bender2002PRL}{C. M. Bender, D. C. Brody, and H. F. Jones, Phys. Rev. Lett. \textbf{89,} 270401 (2002).}

\bibitem{Bender2007RPP}{C. M. Bender, Rep. Prog. Phys. \textbf{70,} 947 (2007).}

\bibitem{Mostafazadeh2002JMP}{A. Mostafazadeh, J. Math. Phys. \textbf{43,} 205 (2002).}

\bibitem{Mostafazadeh2010IJMMP}{A. Mostafazadeh, Int. J. Geom. Meth. Mod. Phys. \textbf{7,} 1191 (2010).}

\bibitem{Moiseyev2011Book}{N. Moiseyev, \emph{Non-Hermitian Quantum Mechanics} (Cambridge University Press, Cambridge, UK, 2011).}

\bibitem{Zeng2020PRB1}{Q.-B. Zeng, Y.-B. Yang, and R. L\"u, Phys. Rev. B \textbf{101,} 125418 (2020).}

\bibitem{Kawabata2020PRR}{K. Kawabata and M. Sato, Phys. Rev. Research \textbf{2,} 033391 (2020).}

\bibitem{Zeng2021arxiv}{Q.-B. Zeng and R. L\"u, New J. Phys. \textbf{24,} 043023 (2022).}

\bibitem{Heiss2012JPAMT}{W. D. Heiss, J. Phys. A: Math. Theor. \textbf{45,} 444016 (2012).}

\bibitem{Xu2017PRL}{Y. Xu, S.-T. Wang, and L.-M. Duan, Phys. Rev. Lett. \textbf{118,} 045701 (2017).}

\bibitem{Gong2018PRX}{Z. Gong, Y. Ashida, K. Kawabata, K. Takasan, S. Higashikawa, and M. Ueda, Phys. Rev. X \textbf{8,} 031079 (2018).}

\bibitem{Carlstrom2018PRA}{J. Carlström and E. J. Bergholtz, Phys. Rev. A \textbf{98,} 042114 (2018).}

\bibitem{Carlstrom2019PRB}{J. Carlström, M. Stålhammar, J. C. Budich, and E. J. Bergholtz, Phys. Rev. B \textbf{99,} 161115(R) (2019).}

\bibitem{Yang2019PRB}{Z. Yang and J. Hu, Phys. Rev. B \textbf{99,} 081102(R) (2019).}

\bibitem{Yang2020PRL1}{Z. Yang, C.-K. Chiu, C. Fang, and J. Hu, Phys. Rev. Lett. \textbf{124,} 186402 (2020).}

\bibitem{Hu2021PRL}{H. Hu and E. Zhao, Phys. Rev. Lett. \textbf{126,} 010401 (2021).}

\bibitem{Yang2021CPL}{X. M. Yang, L. Jin, and Z. Song, Chinese Phys. Lett. \textbf{38,} 060302 (2021).}

\bibitem{Lee2016PRL}{T. E. Lee, Phys. Rev. Lett. \textbf{116,} 133903 (2016).}

\bibitem{Lieu2018PRB}{S. Lieu, Phys. Rev. B \textbf{97,} 045106 (2018).}

\bibitem{Yin2018PRA}{C. Yin, H. Jiang, L. Li, R. L\"u, and S. Chen, Phys. Rev. A \textbf{97,} 052115 (2018).}

\bibitem{Xiong2018JPC}{Y. Xiong, J. Phys. Commun. \textbf{2,} 035043 (2018).}

\bibitem{Shen2018PRL}{H. Shen, B. Zhen, and L. Fu, Phys. Rev. Lett. \textbf{120,} 146402 (2018).}

\bibitem{Yao2018PRL1}{S. Yao and Z. Wang, Phys. Rev. Lett. \textbf{121,} 086803 (2018).}

\bibitem{Yao2018PRL2}{S. Yao, F. Song, and Z. Wang, Phys. Rev. Lett. \textbf{121,} 136802 (2018).}

\bibitem{Alvarez2018PRB}{V. M. Martinez Alvarez, J. E. Barrios Vargas, and L. E. F. Foa Torres, Phys. Rev. B \textbf{97,} 121401(R) (2018).}

\bibitem{Alvarez2018EPJ}{V. M. Martinez Alvarez, J. E. Barrios Vargas, M. Berdakin, and L. E. F. Foa Torres, Eur. Phys. J. Spec. Top. \textbf{227,} 1295 (2018).}

\bibitem{Lee2019PRB}{C. H. Lee and R. Thomale, Phys. Rev. B \textbf{99,} 201103(R) (2019).}

\bibitem{Zhou2019PRB}{H. Zhou and J. Y. Lee, Phys. Rev. B \textbf{99,} 235112 (2019).}

\bibitem{Kawabata2019PRX}{K. Kawabata, K. Shiozaki, M. Ueda, and M. Sato, Phys. Rev. X \textbf{9,} 041015 (2019).}

\bibitem{Okuma2020PRB}{N. Okuma and Masatoshi Sato, Phys. Rev. B \textbf{102,} 014203 (2020).}

\bibitem{Xiao2020NatPhys}{L. Xiao, T. Deng, K. Wang, G. Zhu, Z. Wang, W. Yi, and P. Xue, Nat. Phys. \textbf{16,} 761 (2020).}

\bibitem{Yoshida2020PRR}{T. Yoshida, T. Mizoguchi, and Y. Hatsugai,  Phys. Rev. Research \textbf{2,} 022062(R) (2020).}

\bibitem{Longhi2019PRR}{S. Longhi, Phys. Rev. Research \textbf{1,} 023013 (2019).}

\bibitem{Yi2020PRL}{Y. Yi and Z. Yang, Phys. Rev. Lett. \textbf{125,} 186802 (2020).}

\bibitem{Kunst2018PRL}{F. K. Kunst, E. Edvardsson, J. C. Budich, and E. J. Bergholtz, Phys. Rev. Lett. \textbf{121,} 026808 (2018).}

\bibitem{Jin2019PRB}{L. Jin and Z. Song, Phys. Rev. B \textbf{99,} 081103(R) (2019).}

\bibitem{Yokomizo2019PRL}{K. Yokomizo and S. Murakami, Phys. Rev. Lett. \textbf{123,} 066404 (2019).}

\bibitem{Herviou2019PRA}{L. Herviou, J. H. Bardarson, and N. Regnault, Phys. Rev. A \textbf{99,} 052118 (2019).}

\bibitem{Yang2020PRL2}{Z. Yang, K. Zhang, C. Fang, and J. Hu, Phys. Rev. Lett. \textbf{125,} 226402 (2020).}

\bibitem{Zirnstein2021PRL}{H.-G. Zirnstein, G. Refael, and B. Rosenow, Phys. Rev. Lett. \textbf{126,} 216407 (2021).}

\bibitem{Zhang2022arxiv}{Z. Q. Zhang, H. Liu, H. Liu, H. Jiang, and X. C. Xie, arXiv:2201.01577 (2022).}

\bibitem{Hatano1996PRL}{N. Hatano and D. R. Nelson, Phys. Rev. Lett. \textbf{77,} 570 (1996).}

\bibitem{Shnerb1998PRL}{N. M. Shnerb and D. R. Nelson, Phys. Rev. Lett. \textbf{80,} 5172 (1998).}

\bibitem{Jiang2019PRB}{ H. Jiang, L.-J. Lang, C. Yang, S.-L. Zhu, and S. Chen, Phys. Rev. B \textbf{100,} 054301 (2019).}

\bibitem{Zeng2020PRR}{Q.-B. Zeng and Y. Xu, Phys. Rev. Research \textbf{2,} 033052 (2020).}

\bibitem{Liu2021PRB1}{Y. Liu, Y. Wang, X. J. Liu, Q. Zhou, and S. Chen, Phys. Rev. B \textbf{103,} 014203 (2021).}

\bibitem{Liu2021PRB2}{Y. Liu, Q. Zhou, and S. Chen, Phys. Rev. B \textbf{104,} 024201 (2021).}

\bibitem{Borgnia2020PRL}{D. S. Borgnia, A. J. Kruchkov, and R.-J. Slager, Phys. Rev. Lett. \textbf{124,} 056802 (2020).}

\bibitem{Okuma2020PRL}{N. Okuma, K. Kawabata, K. Shiozaki, and M. Sato, Phys. Rev. Lett. \textbf{124,} 086801 (2020).}

\bibitem{Zhang2020PRL}{K. Zhang, Z. Yang, and C. Fang, Phys. Rev. Lett. \textbf{125,} 126402 (2020).}

\bibitem{Dias2022PRB}{R. G. Dias and A. M. Marques, Phys. Rev. B \textbf{105,} 035102 (2022).}

\bibitem{Shen2021arxiv}{R. Shen and C. H. Lee, arXiv:2107.03414 (2021).}

\bibitem{Imhof2018NatPhys}{S. Imhof, C. Berger, F. Bayer, J. Brehm, L. W. Molenkamp, T. Kiessling, F. Schindler, C. H. Lee, M. Greiter, T. Neupert, and R. Thomale, Nat. Phys. \textbf{14,} 925 (2018).}

\bibitem{Hofmann2019PRL}{T. Hofmann, T. Helbig, C. H. Lee, M. Greiter, and R. Thomale, Phys. Rev. Lett. \textbf{122,} 247702 (2019).}

\bibitem{Helbig2019arxiv}{H. Helbig, T. Hofmann, S. Imhof, M. Abdelghany, T. Kiessling, L. W. Molenkamp, C. H. Lee, A. Szameit, M. Greiter, and R. Thomale, arXiv:1907.11562.}

\bibitem{Chen2009}{W.-K. Chen, \emph{The Circuits and Filters Handbook}, 3rd ed. (CRC, Boca Raton, FL, 2009).}

\end{thebibliography}
\end{document}